\documentclass[%
 reprint,
 superscriptaddress,
 showpacs, preprintnumbers,
 nofootinbib,
 amsmath, amssymb,
 aps,
 prc,
]{revtex4-1}

\usepackage{graphicx}% Include figure files
\usepackage{dcolumn}% Align table columns on decimal point
\usepackage{bm}% bold math
\begin{document}

%\preprint{APS/123-QED}

\title{High-resolution search for the $\Theta^{+}$ pentaquark via a pion-induced reaction at J-PARC}

\newcommand{\KyotoU}{Department of Physics, Kyoto University, Kyoto 606-8502, Japan}
\newcommand{\TorinoTec}{Dipartimento di Scienza Applicata e Tecnologia, Politecnico di Torino, I-10129 Torino, Italy}
\newcommand{\INFN}{INFN, Istituto Nazionale di Fisica Nucleare, Sez. di Torino, I-10125 Torino, Italy}
\newcommand{\RCNP}{Research Center for Nuclear Physics (RCNP), Osaka University, Ibaraki 567-0047, Japan}
\newcommand{\KEK}{High Energy Accelerator Research Organization (KEK), Tsukuba 305-0801, Japan}
\newcommand{\SNU}{Department of Physics and Astronomy, Seoul National University, Seoul 151-747, Republic of Korea}
\newcommand{\NewMexico}{Department of Physics and Astronomy, University of New Mexico, NM 87131-0001, USA}
\newcommand{\TorinoU}{Dipartimento di Fisica, Universit$\grave{a}$ di Torino, I-10125 Torino, Italy}
\newcommand{\TohokuU}{Department of Physics, Tohoku University, Sendai 980-8578, Japan}
\newcommand{\Dubna}{Joint Institute for Nuclear Research, Dubna, Moscow Region 141980, Russia}
\newcommand{\OsakaU}{Department of Physics, Osaka University, Toyonaka 560-0043, Japan}
\newcommand{\JAEA}{Japan Atomic Energy Agency (JAEA), Tokai, Ibaraki 319-1195, Japan}
\newcommand{\RIKEN}{RIKEN, Wako, Saitama 351-0198, Japan}
\newcommand{\TokyoU}{Department of Physics, University of Tokyo, Tokyo 113-0033, Japan}
\newcommand{\ITEP}{ITEP, Institute of Theoretical and Experimental Physics, Moscow 117218, Russia}

\author{M.~Moritsu}  \email{moritsu@rcnp.osaka-u.ac.jp}  \altaffiliation[Present address: ]{\RCNP}  \affiliation{\KyotoU}
\author{S.~Adachi}  \affiliation{\KyotoU}
\author{M.~Agnello}  \affiliation{\TorinoTec}  \affiliation{\INFN}
\author{S.~Ajimura}  \affiliation{\RCNP}
\author{K.~Aoki}  \affiliation{\KEK}
\author{H.~C.~Bhang}  \affiliation{\SNU}
\author{B.~Bassalleck}  \affiliation{\NewMexico}
\author{E.~Botta}  \affiliation{\TorinoU}  \affiliation{\INFN}
\author{S.~Bufalino}  \affiliation{\INFN}
\author{N.~Chiga}  \affiliation{\TohokuU}
\author{H.~Ekawa}  \affiliation{\KyotoU}
\author{P.~Evtoukhovitch}  \affiliation{\Dubna}
\author{A.~Feliciello}  \affiliation{\INFN}
\author{H.~Fujioka}  \affiliation{\KyotoU}
\author{S.~Hayakawa}  \affiliation{\OsakaU}
\author{F.~Hiruma}  \affiliation{\TohokuU}
\author{R.~Honda}  \affiliation{\TohokuU}
\author{K.~Hosomi}  \altaffiliation[Present address: ]{\JAEA}  \affiliation{\TohokuU}
\author{Y.~Ichikawa}  \affiliation{\KyotoU}
\author{M.~Ieiri}  \affiliation{\KEK}
\author{Y.~Igarashi}  \affiliation{\KEK}
\author{K.~Imai}  \affiliation{\JAEA}
\author{N.~Ishibashi}  \affiliation{\OsakaU}
\author{S.~Ishimoto}  \affiliation{\KEK}
\author{K.~Itahashi}  \affiliation{\RIKEN}
\author{R.~Iwasaki}  \affiliation{\KEK}
\author{C.~W.~Joo}  \affiliation{\SNU}
\author{S.~Kanatsuki}  \affiliation{\KyotoU}
\author{M.~J.~Kim}  \affiliation{\SNU}
\author{S.~J.~Kim}  \affiliation{\SNU}
\author{R.~Kiuchi}  \altaffiliation[Present address: ]{\KEK}  \affiliation{\SNU}
\author{T.~Koike}  \affiliation{\TohokuU}
\author{Y.~Komatsu}  \affiliation{\TokyoU}
\author{V.~V.~Kulikov}  \affiliation{\ITEP}
\author{S.~Marcello}  \affiliation{\TorinoU}  \affiliation{\INFN}
\author{S.~Masumoto}  \affiliation{\TokyoU}
\author{Y.~Matsumoto}  \affiliation{\TohokuU}
\author{K.~Matsuoka}  \affiliation{\OsakaU}
\author{K.~Miwa}  \affiliation{\TohokuU}
\author{T.~Nagae}  \affiliation{\KyotoU}
\author{M.~Naruki}  \altaffiliation[Present address: ]{\KyotoU}  \affiliation{\KEK}
\author{M.~Niiyama}  \affiliation{\KyotoU}
\author{H.~Noumi}  \affiliation{\RCNP}
\author{Y.~Nozawa}  \affiliation{\KyotoU}
\author{R.~Ota}  \affiliation{\OsakaU}
\author{K.~Ozawa}  \affiliation{\KEK}
\author{N.~Saito}  \affiliation{\KEK}
\author{A.~Sakaguchi}  \affiliation{\OsakaU}
\author{H.~Sako}  \affiliation{\JAEA}
\author{V.~Samoilov}  \affiliation{\Dubna}
\author{M.~Sato}  \affiliation{\TohokuU}
\author{S.~Sato}  \affiliation{\JAEA}
\author{Y.~Sato}  \affiliation{\KEK}
\author{S.~Sawada}  \affiliation{\KEK}
\author{M.~Sekimoto}  \affiliation{\KEK}
\author{K.~Shirotori}  \altaffiliation[Present address: ]{\RCNP}  \affiliation{\TohokuU}  \affiliation{\JAEA}
\author{H.~Sugimura}  \altaffiliation[Present address: ]{\JAEA}  \affiliation{\KyotoU}
\author{S.~Suzuki}  \affiliation{\KEK}
\author{H.~Takahashi}  \affiliation{\KEK}
\author{T.~Takahashi}  \affiliation{\KEK}
\author{T.~N.~Takahashi}  \altaffiliation[Present address: ]{\RCNP}  \affiliation{\RIKEN}  \affiliation{\TokyoU}
\author{H.~Tamura}  \affiliation{\TohokuU}
\author{T.~Tanaka}  \affiliation{\OsakaU}
\author{K.~Tanida}  \affiliation{\SNU}  \affiliation{\JAEA}
\author{A.~O.~Tokiyasu}  \altaffiliation[Present address: ]{\RCNP}  \affiliation{\KyotoU}
\author{N.~Tomida}  \affiliation{\KyotoU}
\author{Z.~Tsamalaidze}  \affiliation{\Dubna}
\author{M.~Ukai}  \affiliation{\TohokuU}
\author{K.~Yagi}  \affiliation{\TohokuU}
\author{T.~O.~Yamamoto}  \affiliation{\TohokuU}
\author{S.~B.~Yang}  \affiliation{\SNU}
\author{Y.~Yonemoto}  \affiliation{\TohokuU}
\author{C.~J.~Yoon}  \affiliation{\SNU}
\author{K.~Yoshida}  \affiliation{\OsakaU}

\collaboration{J-PARC E19 Collaboration}  \noaffiliation

\date{22 September 2014}

%%%%%%%%%%%%%%%%%%%%%%%%%%%%%%%%%%%%%%%%%%%%%%%%%%%%%%%%%%%%%%%%%%%%%%%%%%%%%%%%%%%%%%
\begin{abstract}
The pentaquark $\Theta^+$ has been searched for via the $\pi^-p \to K^-X$ reaction with beam momenta of 1.92 and 2.01 GeV/$c$ at J-PARC. A missing mass resolution of 2 MeV (FWHM) was achieved but no sharp peak structure was observed. The upper limits on the production cross section averaged over the scattering angle from 2$^{\circ}$ to 15$^{\circ}$ in the laboratory frame were found to be less than 0.28 $\mu$b/sr at the 90\% confidence level for both the 1.92- and 2.01-GeV/$c$ data. The systematic uncertainty of the upper limits was controlled within 10\%. Constraints on the $\Theta^+$ decay width were also evaluated with a theoretical calculation using effective Lagrangian. The present result implies that the width should be less than 0.36 and 1.9 MeV for the spin-parity of $1/2^+$ and $1/2^-$, respectively. 
\end{abstract}

%% PACS, the Physics and Astronomy Classification Scheme.
  %Pion-induced reaction

%\keywords{Suggested keywords}%Use showkeys class option if keyword display desired

\maketitle
%\tableofcontents

%%%%%%%%%%%%%%%%%%%%%%%%%%%%%%%%%%%%%%%%%%%%%%%%%%%%%%%%%%%%%%%%%%%%%%%%%%%%%%%%%%%%%%
%%%                                   INTRODUCTION
%%%%%%%%%%%%%%%%%%%%%%%%%%%%%%%%%%%%%%%%%%%%%%%%%%%%%%%%%%%%%%%%%%%%%%%%%%%%%%%%%%%%%%
\section{\label{sec:intro}Introduction}

Study of exotic hadrons, which cannot be interpreted as ordinary three-quark baryons or quark-antiquark mesons, has a long history starting in the 1970s. An exotic pentaquark $\Theta^+(1540)$ has received enthusiastic attention and numerous papers have been published since the first evidence was reported by the LEPS Collaboration in 2003 \cite{Nakano2003}. The $\Theta^+$ baryon has a strangeness quantum number $S=+1$ with its minimal quark configuration of $uudd\bar{s}$. A possible existence of $\Theta^+$ was first advocated by Diakonov, Petrov, and Polyakov using a chiral soliton model \cite{Diakonov1997}. They predicted an exotic positive-strangeness baryon, having spin-parity $1/2^+$ and isospin 0, with a light mass of about 1530 MeV/$c^2$ and a width of less than 15 MeV. 

Soon after the first experimental evidence, several experimental groups published supporting evidence for $\Theta^+$, followed by a number of experiments with no evidence; see Refs.~\cite{Hicks2005,Hicks2012} for reviews. The experimental situation became controversial. The mass of $\Theta^+$ claimed in each experiment ranged from 1520 to 1550 MeV/$c^2$. Besides, the spin and parity have not been determined yet experimentally. 

In the $\gamma d \to K^+K^-pn$ reaction, LEPS confirmed their original evidence \cite{Nakano2009}. A peak was observed in the Fermi-motion-corrected $nK^+$ invariant mass distribution with the statistical significance of $5.1\sigma$. In their recent preliminary result with increased statistics, the significance decreased but an enhancement remained in data enriching the quasifree $\gamma n$ reaction \cite{Kato2013}. In contrast, the CLAS Collaboration also searched for the $\Theta^+$ in the same reaction but observed no peak \cite{McKinnon2006}. Although these results seem to be inconsistent, the discrepancy might be attributed to the different angular acceptances of these two experiments. If the $\Theta^+$ production cross section has a strong angular dependence peaked forward, only the LEPS detector will show a signal. An alternative theoretical explanation for the LEPS result was proposed by Mart\'inez Torres and Oset \cite{Torres2010}. Based on their $\gamma d \to K^+K^-pn$ reaction calculation without $\Theta^+$ production, they claimed that the statistical significance was only $2\sigma$. 

Recently, Amaryan {\it et al.} reported an observation of a narrow peak in the missing mass of $K^0_S$ in the $\gamma p \to p K^{0}_{S}K^{0}_{L}$ reaction using data from the CLAS detector \cite{Amaryan2012}. They claimed that the peak may be due to the interference between the $\Theta^+$ and $\phi$ leading to the same final state, whereas the CLAS Collaboration itself was not convinced of the evidence \cite{Anghinolfi2012}. 

The formation reaction, $K^+n \to \Theta^+$, which is a reverse reaction of the decay, was investigated by the DIANA and Belle Collaborations. In this reaction, the $\Theta^+$ width, $\Gamma_{\Theta}$, can be derived from the cross section. Belle observed no evidence using kaon secondary interactions in the detector materials and set an upper limit of $\Gamma_{\Theta}<0.64$ MeV at the 90\% C.L. \cite{Mizuk2006}. On the other hand, DIANA observed evidence for the $\Theta^+$ in the $pK^{0}_{S}$ invariant mass spectra from the $K^+{\rm Xe} \to K^0p{\rm Xe'}$ reaction \cite{Diana,Diana2014}. The width was estimated to be $0.34\pm0.10$ MeV. 

A narrow width is peculiar to the $\Theta^+$. According to reanalyses of old $K^+d$ scattering data \cite{Arndt2003,Haidenbauer2003,Cahn2004,Sibirtsev2004,Gibbs2004}, there is a consensus that the width should be less than a few MeV if the $\Theta^+$ exists. It is quite narrower than those of ordinary strongly decaying hadrons. The problem of the narrow width is coupled with the structure of $\Theta^+$. In naive consideration, an $s$-wave resonance having negative parity is unlikely. Jaffe-Wilczek \cite{Jaffe2003} and Karliner-Lipkin \cite{Karliner2003} proposed a $\Theta^+$ internal structure based on diquark correlation. Both predicted positive parity with finite relative angular momentum. They suggested that the narrow width might be explained with rearrangement of the color, spin, and spatial wave functions needed in the $\Theta^+ \to KN$ decay. In this manner, the narrowness of $\Theta^+$ is strongly related to the low-energy quark dynamics. 

So far, searches using meson-induced reactions were performed via the $\pi^-p \to K^-X$ and $K^+p \to \pi^+X$ reactions in the KEK E522 \cite{Miwa2006} and E559 \cite{Miwa2008} experiments, respectively. No significant peak was observed in either reaction; however, a bump structure at 1530 MeV/$c^2$ was reported in the $\pi^-p \to K^-X$ reaction at a beam momentum of 1.92 GeV/$c$. Since the statistical significance was only 2.5--2.7$\sigma$, they did not claim evidence and derived an upper limit of the forward production cross section of 2.9 $\mu$b/sr at the 90\% C.L. Their significance was limited by the poor mass resolution of 13 MeV (FWHM). We can easily improve the resolution by an order of magnitude with a good spectrometer system. 

Under the current situation, the present experimental search should satisfy the following requirements. (i) High-statistics data are indispensable in order not to be disturbed by statistical fluctuation. (ii) High resolution of less than a few MeV is desirable to measure the potentially narrow $\Theta^+$. 

We intended to investigate the $\Theta^+$ using the $\pi^-p \to K^-X$ reaction. It was timely to use high-intensity meson beams at the recently constructed J-PARC facility \cite{JPARC}. We have constructed a high-resolution spectrometer system in order to achieve a good mass resolution of 2 MeV (FWHM). Since we used the missing mass technique with a liquid hydrogen target, we could avoid corrections for the Fermi motion or rescattering effect. An order of magnitude higher sensitivity than the previous E522 experiment was expected. 

In this paper, we present the results of a search for the $\Theta^+$ via the $\pi^-p \to K^-X$ reaction. The result at 1.92 GeV/$c$ momentum was reported in a previous Letter \cite{Shirotori2012}. This paper reports the results at 2.01 GeV/$c$ momentum including details of the experimental apparatus and the analysis procedures. A discussion of the $\Theta^+$ width based on the present results with a theoretical calculation is also given.

%%%%%%%%%%%%%%%%%%%%%%%%%%%%%%%%%%%%%%%%%%%%%%%%%%%%%%%%%%%%%%%%%%%%%%%%%%%%%%%%%%%%%%
%%%                                    EXPERIMENT
%%%%%%%%%%%%%%%%%%%%%%%%%%%%%%%%%%%%%%%%%%%%%%%%%%%%%%%%%%%%%%%%%%%%%%%%%%%%%%%%%%%%%%
\section{\label{sec:exp}Experiment}

\begin{figure}[tb] %------------------------------------------------------
\includegraphics[width=8cm]{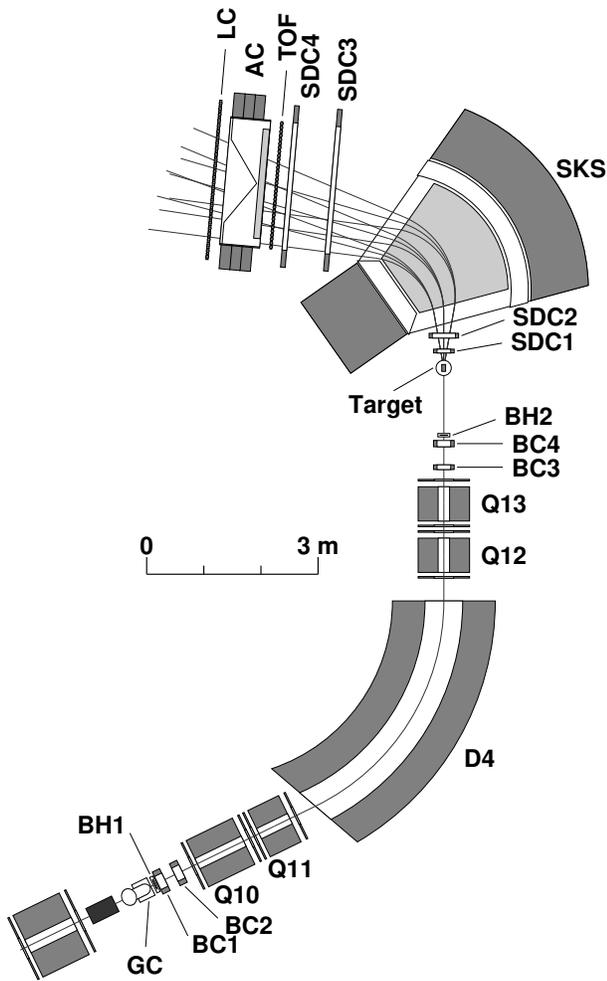}
\caption{\label{fig:setup} Schematic view of the experimental setup in 2012.}
\end{figure}

We have performed the experiment (J-PARC E19) which is a high-resolution search for the $\Theta^+$ pentaquark via the $\pi^-p \to K^-X$ reaction. In order to realize a high-resolution missing mass spectroscopy, we have constructed two spectrometers \cite{Takahashi2012}: the beam spectrometer and the superconducting kaon spectrometer (SKS). Figure~\ref{fig:setup} shows a schematic view of the experimental setup. Physics data were taken in 2010 and 2012 using different beam momenta of 1.92 and 2.01 GeV/$c$, respectively. The first run \cite{Shirotori2012} was carried out to have a direct comparison with the previous E522 experimental result. The second run was performed using the maximum beam momentum of the K1.8 beam line. We chose the higher momentum because an increase in the production cross section was expected from a theoretical prediction \cite{Oh2004-1,Hyodo2012}.

 %----------------------------------------------------------

%%%%%%%%%%%%%%%%%%%%%%%%%%%%%%%%%%%%%%%%%%%%%%%%%%%%%%%%%%
\subsection{\label{ssec:k18}K1.8 beam line}

A primary proton beam was extracted from the J-PARC 30-GeV proton synchrotron for 2.2-s spill in 6-s repetition to a platinum production target. The K1.8 beam line \cite{Agari2012} is a separated secondary-particle beam line up to 2 GeV/$c$ momentum. The beam line has two electrostatic separators which were designed to obtain high-purity kaon beams. A typical beam size was 10 (horizontal) $\times$ 5 (vertical) mm$^2$ (rms) at the experimental target. The central beam momenta were set at 1.92 or 2.01 GeV/$c$ with a spread of typically 1\% rms. The average beam intensity was adjusted to 1.0 and 1.7 $\times 10^6$/spill in 2010 and 2012, respectively, which was limited by an acceptable instantaneous rate. Due to the beam ripples, the maximum instantaneous rate became up to twenty times as high as the mean rate \cite{Takahashi2012}. \footnote{The acceptable beam intensity was increased by ten times in 2013 thanks to efforts by the accelerator group and a detector upgrade \cite{Sugimura2014}.}

%%%%%%%%%%%%%%%%%%%%%%%%%%%%%%%%%%%%%%%%%%%%%%%%%%%%%%%%%%
\subsection{\label{ssec:bs}Beam spectrometer}

The last part of the K1.8 beam line is the beam spectrometer \cite{Takahashi2012}. It comprises a $QQDQQ$ magnet system with four sets of wire chambers (BC1--BC4), a gas Cherenkov counter (GC), and two sets of segmented plastic scintillation counters (BH1 and BH2).

BC1 and BC2 were 1-mm pitch multiwire proportional chambers (MWPCs) installed at the upstream part of the $QQDQQ$ system. At the exit of the last $Q$-magnet, drift chambers BC3 and BC4 which have drift spaces of 1.5 and 2.5 mm, respectively, were installed. Beam tracks were measured with a position resolution of 200 $\mu$m. Beam momenta were reconstructed particle by particle with a resolution of $10^{-3}$ (FWHM). In order to minimize the multiple-scattering effect on the momentum resolution, the beam spectrometer optics was designed to realize point-to-point focus to the first order. The magnetic field of the dipole magnet was monitored during the experimental period by a high-precision Hall probe.

BH1 and BH2 were used as trigger and time-of-flight counters for beam particles with a time resolution of 0.2 ns. GC is a pressured isobutane gas Cherenkov counter (n = 1.002). It vetoed electrons, which contaminated 10--20\% of the beam, with a rejection efficiency of 99.5\%. The beam trigger was defined as BEAM $\equiv$ BH1 $\times$ BH2 $\times$ $\overline{\rm GC}$.

\begin{table*}[tb] %-----------------------------------------------------------
\caption{\label{tab:runsum} Experimental data summary. The first row is the $\Theta^+$ search data. Others are calibration data.}
\begin{ruledtabular}
\begin{tabular}{ccccc}
 Reaction & Beam momentum (GeV/$c$) & Target & \multicolumn{2}{c}{Number of particles in beam} \\ \cline{4-5}
 & & & 2010 data & 2012 data \\ \hline
 $\pi^-p \to K^-X$ & 1.92 / 2.01 & LH$_2$ & $7.8 \times 10^{10}$ & $8.1 \times 10^{10}$ \\
 Empty run \footnote{$(\pi^-,\pi^-)$ scattering events were used for analysis.} 
                          & 1.92 / 2.01 & empty   & $4.6 \times 10^{9}$  & $4.1 \times 10^{9}$ \\
 $\pi^+p \to K^+\Sigma^+$ & 1.38 & LH$_2$ & $2.9 \times 10^{9}$ & $8.5 \times 10^{8}$ \\
 $\pi^-p \to K^+\Sigma^-$ & 1.38 & LH$_2$ & $1.2 \times 10^{10}$ & $3.8 \times 10^{9}$ \\
 $\pi^-p \to K^+\Sigma^-$ & 1.46 & LH$_2$ &     & $8.7 \times 10^{9}$ \\
 $\pi^{\pm}$ beam-through & 0.75--1.38 & empty &     &    \\ 
\end{tabular}
\end{ruledtabular}
\end{table*} %-----------------------------------------------------------

%%%%%%%%%%%%%%%%%%%%%%%%%%%%%%%%%%%%%%%%%%%%%%%%%%%%%%%%%%
\subsection{\label{ssec:sks}SKS spectrometer}

The SKS spectrometer \cite{Takahashi2012} comprises a superconducting dipole magnet with four sets of drift chambers (SDC1--SDC4) and three kinds of trigger counters (TOF, AC, and LC). It establishes both a good momentum resolution of $2 \times 10^{-3}$ (FWHM) and a large acceptance of 100 msr around 1-GeV/$c$ momentum. The SKS magnet which had been utilized in the KEK-PS experiments was moved to the J-PARC hadron facility. Details of the original specification are described elsewhere \cite{SKS1995}. 

SDC1 and SDC2 were installed at the entrance of the magnet, which have the same drift-cell structure as that of BC3 and BC4. Large-area drift chambers, SDC3 and SDC4, were placed at the exit of the magnet. Since the setup of SKS was slightly changed between 2010 and 2012, the momentum acceptance was somewhat different: 0.7--1.0 and 0.8--1.2 GeV/$c$ in 2010 and 2012, respectively. The magnet was excited at 2.5 T and its field was monitored by an NMR probe during the experimental period. Helium bags were installed in the gap of the magnet and between the magnet and SDC3 to avoid multiple scatterings by air. 

The TOF wall consists of 32 vertical plastic scintillation counters performing the time-of-flight measurement for scattered particle identification. AC is a threshold-type Cherenkov counter with silica aerogel (n = 1.05) as the radiator for pion veto. Two small-size ACs were used in the 2010 run, while we replaced them with a new larger-size AC before the 2012 run. The LC wall consists of 28 vertical threshold-type lucite Cherenkov counters (n = 1.49). It was used to discriminate low-momentum protons from pions and kaons. We adopted a matrix-coincidence trigger (MATRIX) considering a hit-segment combination of TOF with LC. It distinguishes reaction events at the target from fake triggers originated from the beam hitting the magnet and detector frames. Production data were triggered by the $(\pi, K)$ reaction events which are defined as PIK $\equiv$ BEAM $\times$ TOF $\times$ $\overline{\rm AC}$ $\times$ LC $\times$ MATRIX. The typical trigger rate was 400 per spill.

%%%%%%%%%%%%%%%%%%%%%%%%%%%%%%%%%%%%%%%%%%%%%%%%%%%%%%%%%%
\subsection{\label{ssec:daq}Experimental target and data acquisition}

We used a liquid hydrogen (LH$_2$) target with a thickness of 0.85 g/cm$^2$. The target size was 67.8 mm in diameter and 120 mm in length along the beam direction. Both end caps of the target vessel and the windows of the target vacuum chamber were made of a 0.25-mm-thick mylar. The stability was monitored during the experimental period and the density fluctuation was less than $3\times10^{-5}$. 

We developed a network-based data acquisition system \cite{Igarashi2010} for various experiments in the hadron facility. It integrated several kinds of readout subsystems through network. In order to validate the data structure among different electronics standards, a trigger/tag distribution system was developed. The tags embedded in the data structure were checked at the beginning of the data decoding.

%%%%%%%%%%%%%%%%%%%%%%%%%%%%%%%%%%%%%%%%%%%%%%%%%%%%%%%%%%
\subsection{\label{ssec:data}Data summary}

Table~\ref{tab:runsum} shows the data summary. For the $\Theta^+$ search data, the $\pi^-p \to K^-X$ reactions at 1.92 and 2.01 GeV/$c$ were accumulated in the 2010 and 2012 runs, respectively. Empty target data with the empty vessel instead of the liquid hydrogen target were also taken in order to estimate a background contamination from surrounding materials and the vertex cut efficiency. For the momentum calibration of the spectrometers, the following two kinds of calibration data were used. 

(i) In the $\pi^{\pm} p \to K^+ \Sigma^{\pm}$ reactions, incident momenta of 1.38 and 1.46 GeV/$c$ were selected so that the central momenta of scattered kaons were 0.9 and 1.0 GeV/$c$, respectively. These kaon momenta were compatible with those generated by the $\Theta^+$ production at 1.92 or 2.01 GeV/$c$ beam momentum. The validity of the analysis was examined by using the missing-mass peak and the cross section of the $\Sigma$ hyperons. 

(ii) A low-momentum pion beam around 1 GeV/$c$ can directly pass through both spectrometers. This kind of data is called $\pi^{\pm}$ beam-through data. The data were acquired at several momenta between 0.75 and 1.38 GeV/$c$ with both positively and negatively charged beams. 
%We define the positive polarity where the positively charged particles are bent into detector acceptance, and vice versa. The $\Theta^+$ search data were taken at the negative polarity of the SKS, whereas the $\Sigma^{\pm}$ production data were taken at the positive polarity. 

%%%%%%%%%%%%%%%%%%%%%%%%%%%%%%%%%%%%%%%%%%%%%%%%%%%%%%%%%%%%%%%%%%%%%%%%%%%%%%%%%%%%%%
%%%                                     ANALYSIS
%%%%%%%%%%%%%%%%%%%%%%%%%%%%%%%%%%%%%%%%%%%%%%%%%%%%%%%%%%%%%%%%%%%%%%%%%%%%%%%%%%%%%%
\section{\label{sec:ana}Analysis}

The $\Theta^+$ was searched for in a missing mass spectrum of the $\pi^-p \to K^-X$ reaction. The missing mass, $M_X$, is calculated in the laboratory frame as follows:
\begin{equation}
M_{X} = \sqrt{(E_{\pi}+m_{p}-E_{K})^2-(p_{\pi}^2+p_{K}^2 -2p_{\pi}p_{K}\cos\theta)},
\label{eq:mm}
\end{equation}
where $E_{\pi}$ and $p_{\pi}$ are the energy and momentum of a beam pion, $E_{K}$ and $p_{K}$ are those of a scattered kaon, $m_p$ is the mass of a target proton, and $\theta$ is the scattering angle defined as the angle between the incoming pion and the outgoing kaon. Thus there are three kinematic variables to be measured; $p_{\pi}, p_K$, and $\theta$. 

The analysis procedure of missing mass reconstruction is described in the next subsection, followed by calibration methods. There is no calibration peak in the $\Theta^+$ search data because the $\pi^-p \to K^-X$ reaction is an exotic channel; therefore, analyses on both the mass scale calibration (Sec.~\ref{ssec:msc}) and the mass resolution (Sec.~\ref{ssec:mr}) were done by using the $\Sigma^{\pm}$ production data and the $\pi^{\pm}$ beam-through data. The cross section calculations are described in Sec.~\ref{ssec:cs}.

\begin{figure}[tb] %-----------------------------------------------------------
\includegraphics[width=\linewidth]{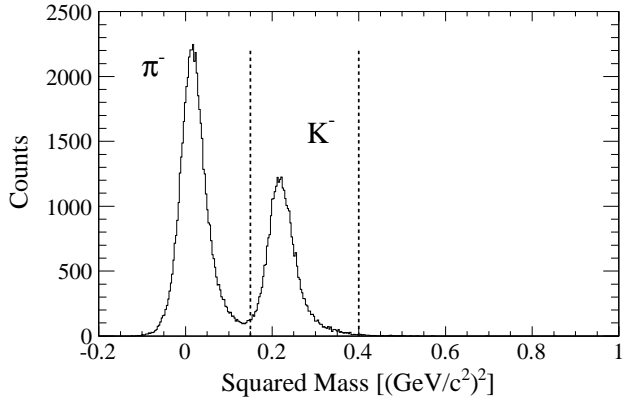}
\caption{\label{fig:pid} Squared mass distribution for scattered particles in the $\Theta^+$ search data in 2012. It is shown in a momentum range of 0.9--1.1 GeV/$c$. The vertex cut and the scattering angle selection of 2--15$^{\circ}$ have been applied. The dashed lines indicate the kaon selection gate.}
\end{figure} %-----------------------------------------------------------

\begin{figure}[tb] %-----------------------------------------------------------
\includegraphics[width=\linewidth]{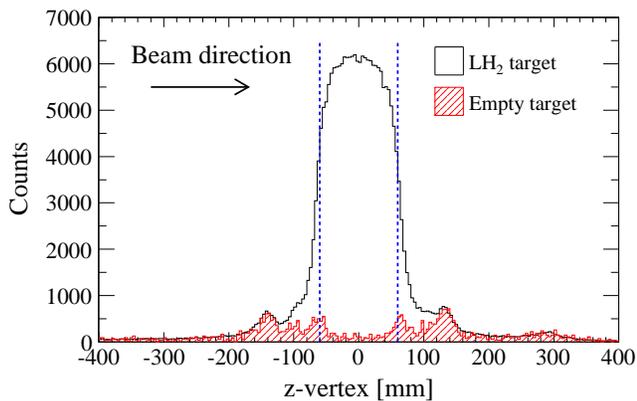}
\caption{\label{fig:vtx}(Color online) Vertex distribution along the $z$ axis (beam direction). The open histogram shows the LH$_2$ target data. The red hatched histogram shows the empty target data normalized by the beam flux. The subtraction of these histograms represents a net contribution from LH$_2$. $(\pi^-, \pi^-)$ events with scattering angles from 2$^{\circ}$ to 15$^{\circ}$ are selected in both histograms. The blue dashed lines indicate the vertex cut positions.}
\end{figure} %-----------------------------------------------------------

%%%%%%%%%%%%%%%%%%%%%%%%%%%%%%%%%%%%%%%%%%%%%%%%%%%%%%%%%%%
\subsection{\label{ssec:mm}Missing mass reconstruction}

The procedure of the analysis was as follows: (i) event selection by using counter information, (ii) momentum reconstruction for beam and scattered particles, (iii) particle identification of kaons, (iv) reconstruction of the scattering angle and the vertex point, and (v) calculation of the missing mass. 

An incident pion was selected by using the time-of-flight information between BH1 and BH2. Then, the beam momentum and the scattered-particle momentum were determined by reconstructing particle trajectories from the hit positions of BCs and SDCs. In the tracking process, straight-line tracks were first defined locally both at the entrance and the exit of each spectrometer by linear least-squares fitting. Then, $\chi^2$ minimization with respect to the momentum vector was done for each combination of the straight-line tracks. In the beam spectrometer, a third-order transport matrix calculated with {\sc orbit} \cite{Morinobu} was utilized for the momentum reconstruction, while, in the SKS spectrometer, trajectories were reconstructed by means of the Runge-Kutta method with a magnetic field map calculated by the finite element method. In the present analysis, events including more than two beam tracks were discarded.  

The mass of a scattered particle was calculated as $M_{\rm scat} = (p/\beta)\sqrt{\mathstrut 1-\beta^2}$, where $\beta$ is the velocity of a scattered particle and $p$ is the momentum determined by the SKS tracking. $\beta$ is calculated with the path length and the time of flight between BH2 and TOF. Figure~\ref{fig:pid} shows a distribution of the squared mass, $M_{\rm scat}^2$, obtained in the $\Theta^+$ search data. The remaining pions are caused by the AC inefficiency of 2\%. The kaon selection cut region is indicated in the spectrum: $0.15<M_{\rm scat}^2<0.40$. Pion contamination in the kaon gate was estimated to be $1.9 \pm 1.0\%$ in a momentum range of 0.9--1.1 GeV/$c$, while the kaon identification efficiency was 96\%. 

The scattering angle and the vertex point were obtained from two tracks: the local straight-line track obtained from BC3 and BC4 and the track obtained by the momentum reconstruction of the SKS. The relative geometry between the beam and SKS spectrometers was adjusted by using the beam-through data. Figure~\ref{fig:vtx} shows a vertex distribution along the $z$ axis (beam direction). Since the $z$-vertex resolution rapidly deteriorates with decrease of the scattering angles, scattering angles less than 2$^{\circ}$ were excluded. This forward angle cut was effective to reject muons which originate from beam pion decay around the target region. In addition, events with scattering angles more than 15$^{\circ}$ were not used in the present analysis because of rapidly decreasing acceptance. The $z$-vertex resolution was estimated to be 10--20 mm for scattering angles from 2$^{\circ}$ to 15$^{\circ}$. In Fig.~\ref{fig:vtx}, contribution from the mylar windows of the target vessel ($z=\pm 60$ mm) and the vacuum chamber ($z=\pm 135$ mm) are clearly seen in the empty target data. The bump around 280 mm is due to SDC1. The vertex cut ($-60 < z < 60$ mm, $\sqrt{\mathstrut x^2+y^2} < 30$ mm) was applied by considering the target vessel size. 

\begin{figure}[tb] %-----------------------------------------------------------
\includegraphics[width=7.5cm]{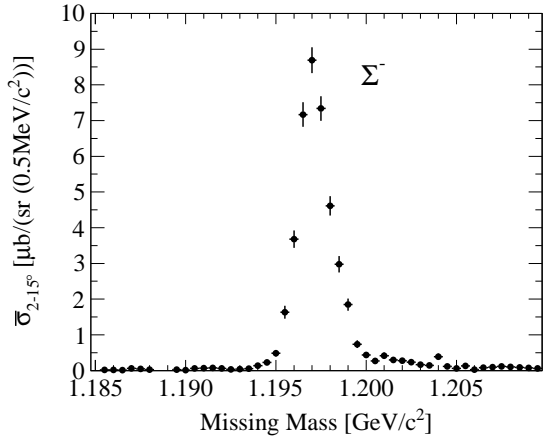}
\caption{\label{fig:sigma1} Missing mass spectrum of the $\pi^-p\to K^+X$ reaction at 1.46 GeV/$c$ after applying the mass scale calibration. The ordinate represents the differential cross section averaged over 2--15$^{\circ}$ in the laboratory frame. The $\Sigma^-$ peak is identified with a resolution of 2.21 $\pm$ 0.05(stat.) $\pm$ 0.1(syst.) MeV in FWHM. The quoted errors in the figure are statistical.}
\end{figure}

Finally the missing mass was calculated according to Eq.~(\ref{eq:mm}). The same analysis procedure was applied to the calibration data. For instance, the missing mass spectrum of the $\pi^-p \to K^+X$ reaction at 1.46 GeV/$c$ is shown in Fig.~\ref{fig:sigma1}, where the $\Sigma^-$ hyperon mass is correctly reconstructed with a resolution of 2.21 $\pm$ 0.05(stat.) $\pm$ 0.1(syst.) MeV (FWHM). The systematic uncertainty was estimated by a fitting range dependence. In the other $\Sigma$ production data, $\Sigma$ hyperon peaks were also reconstructed with similar quality. The obtained peak positions were used in the mass scale calibration, while the peak widths were used to evaluate the $\Theta^+$ mass resolution.

 %-----------------------------------------------------------

%%%%%%%%%%%%%%%%%%%%%%%%%%%%%%%%%%%%%%%%%%%%%%%%%%%%%%%%%%%
\subsection{\label{ssec:msc}Mass scale calibration}

The missing mass scale, in other words, the momentum scales of both spectrometers, was adjusted by several calibration data: $\Sigma^{\pm}$ production data and $\pi^{\pm}$ beam-through data. 

The initial scales of momenta were reconstructed by the beam and SKS spectrometers based on the magnetic-field values monitored by the Hall probe and the NMR probe, respectively. Then, energy loss correction in the LH$_2$ target and the BH2 counter was applied event by event considering the reaction vertex according to the Bethe-Bloch formula. The $\Sigma$ production data provided information on a mass difference between the reconstructed mass and the known $\Sigma$ mass. On the other hand, a momentum difference, $p_{\rm diff} \equiv p_{\rm B}-p_{\rm S}$, was obtained from each beam-through event, where $p_{\rm B}$ ($p_{\rm S}$) is the beam (scattered-particle) momentum at the target position. The momentum scale calibration was done so as to reduce these mass and momentum differences. 

In the momentum scale calibration, we regarded the momentum reconstructed by the SKS as a reference, because the SKS magnet was set at 2.5 T at all times while the beam spectrometer was excited at different field corresponding to the beam momenta. Therefore, we applied a correction on the beam momentum so as to minimize the momentum and mass differences. At first, a polarity offset of 2.7 MeV/$c$ due to the different polarity setting of the beam spectrometer was corrected. Then, a linear correction with respect to the beam momentum was applied. As a result, the momentum differences were reduced within 2 MeV/$c$, and the mass differences were within 1 MeV/$c^2$. We regarded the remaining differences as a systematic uncertainty on the momentum scale, which was estimated to be 0.12\% at most. Considering the momentum scale uncertainty and the kinematics of the $\Theta^+$ production, the missing mass scale uncertainty on $\Theta^+(1530)$ was estimated to be 1.4 MeV/$c^2$ in the 2012 data. The corresponding value in the 2010 data was 1.7 MeV/$c^2$.

%%%%%%%%%%%%%%%%%%%%%%%%%%%%%%%%%%%%%%%%%%%%%%%%%%%%%%%%%%%
\subsection{\label{ssec:mr}Mass resolution}

As for the missing mass spectroscopy, the mass resolution ($\Delta M$) is derived from the momentum resolution of beam and scattered particles ($\Delta p_{\rm B}$ and $\Delta p_{\rm S}$), the scattering angle resolution ($\Delta \theta$) and the energy-loss straggling effect ($\Delta E_{\rm strag}$). It can be expressed in the following equations: 
\begin{eqnarray}
\Delta M^2 &=& \left( \frac{\partial M}{\partial p_{\rm B}}\right) ^2 \Delta p_{\rm B}^2
               + \left( \frac{\partial M}{\partial p_{\rm S}}\right) ^2 \Delta p_{\rm S}^2 \nonumber \\
          && + \left( \frac{\partial M}{\partial \theta}\right) ^2 \Delta \theta ^2
               + \Delta E_{\rm strag}^2 , \label{eq:mresol} \\
\frac{\partial M}{\partial p_{\rm B}} &=& \frac{1}{M} \left[ \beta_{\rm B} (m_p - E_{\rm S})+p_{\rm S} \cos{\theta}\right] , \label{eq:dmdb} \\
\frac{\partial M}{\partial p_{\rm S}} &=& -\frac{1}{M} \left[ \beta_{\rm S} (m_p + E_{\rm B})-p_{\rm B} \cos{\theta}\right] , \label{eq:dmds} \\
\frac{\partial M}{\partial \theta} &=& -\frac{1}{M} p_{\rm B} p_{\rm S} \sin{\theta} . \label{eq:dmdt}
\end{eqnarray}
These are derived from Eq.~(\ref{eq:mm}) where the subscripts $\pi$ and $K$ should be replaced by B and S, which represent beam and scattered particle, respectively, and $\beta$ is the velocity of each particle. The covariance terms in Eq.~(\ref{eq:mresol}) were ignored in the present analysis; since $p_{\rm B}$ and $p_{\rm S}$ are obtained by independent spectrometers, they have no correlation. The correlation between the scattering angle $\theta$ and each momentum was neglected because their contribution to the overall resolution was limited. 

In the previous Letter \cite{Shirotori2012}, we had assumed $\Delta p_{\rm B}/p_{\rm B} = 5.2 \times 10^{-4}$ (FWHM) which was calculated by the first-order transport matrix with a tracker position resolution of 0.2 mm and by fluctuation of the magnetic field. Because the above value might be underestimated, we applied another estimation in the present analysis. As well as the mass scale calibration, the mass resolution for $\Theta^+$ was evaluated by the calibration data: $\Sigma^{\pm}$ production data and $\pi^{\pm}$ beam-through data. 

$\Delta E_{\rm strag}$ denotes contribution from energy-loss straggling in the target and BH2 to the missing mass resolution. It was calculated to be 0.39 MeV for the $\Theta^+$ production at 2 GeV/$c$ assuming the Landau distribution. The dependence on the reaction vertex point in the target was less than 0.01 MeV. 

The stability of the magnetic field was monitored during data acquisition. The long-term fluctuations of the beam and SKS spectrometer field were less than $2.4 \times 10^{-4}$ and $9.6 \times 10^{-5}$, respectively, which were neglected in the momentum resolution. 

The missing mass resolution for the $\Sigma$'s was estimated by fitting the calibration peaks of the $\Sigma$ hyperons. As can be seen in Eqs.~(\ref{eq:mresol}--\ref{eq:dmdt}), the missing mass resolution depends on the scattering angle $\theta$. At forward angles of 2--15$^{\circ}$, the $\theta$ dependence predominantly comes from the $\sin{\theta}$ term in Eq.~(\ref{eq:dmdt}). Hence, $\Delta \theta$ was obtained by fitting the $\theta$ dependence in the $\Sigma$ mass resolution as a function of $\sin{\theta}$. $\Delta \theta$ was estimated to be 5.7 $\pm$ 0.8 mrad (FWHM). 

In order to obtain $\Delta p_{\rm B}$ and $\Delta p_{\rm S}$, another equation was necessary. From the $p_{\rm diff}$ ($=p_{\rm B}-p_{\rm S}$) distribution obtained from the beam-through data, we obtained the width of the distribution, $\Delta p_{\rm diff}$, which is composed of the momentum resolution of the beam and SKS spectrometers and the energy-loss straggling effect in BH2 ($\Delta p_{\rm strag}$). It is written as
\begin{equation}
\Delta p_{\rm diff}^2 = \Delta p_{\rm B}^2 + \Delta p_{\rm S}^2 + \Delta p_{\rm strag}^2 .
\label{eq:presol}
\end{equation}
The typical $\Delta p_{\rm diff}$ value was 2.67 $\pm$ 0.12 MeV/$c$ obtained from the 1.1-GeV/$c$ beam-through data. The effect from different magnet polarity was examined by several sets of beam-through data with the same momentum but with the opposite charge. As a result, the effect on the momentum resolution was negligibly small ($<0.12$ MeV/$c$). $\Delta p_{\rm strag}$ was calculated in the same procedure as for $\Delta E_{\rm strag}$.

Assuming that the momentum resolution is simply proportional to the momentum, $\Delta p_{\rm B}/p_{\rm B}$ and $\Delta p_{\rm S}/p_{\rm S}$ were derived by solving the quadratic equations (\ref{eq:mresol}) and (\ref{eq:presol}). The resolution of the beam spectrometer was calculated to be $\Delta p_{\rm B}/p_{\rm B} = (1.4 \pm 0.2) \times 10^{-3}$ and that of the SKS spectrometer was $\Delta p_{\rm S}/p_{\rm S} = (2.0 \pm 0.2) \times 10^{-3}$ in FWHM. Then, the mass resolution expected for the $\Theta^+$ production was derived as a function of the scattering angle: $\Delta M_{\Theta}(\theta)$. In order to utilize in the following analysis, we needed the mass resolution averaged over 2--15$^{\circ}$, which was found to be 2.13 $\pm$ 0.15 MeV (FWHM) assuming isotropic angular distribution for the $\Theta^+$ production. The dependence on the angular distribution was examined and found to be less than 0.1 MeV because the experimental acceptance was limited to forward angles. The mass resolution in the 2010 data was similarly reevaluated to be 1.72 MeV (FWHM) \cite{Tomonori2014}. The difference between the 2010 and 2012 data was mainly due to the different momentum settings.

%%%%%%%%%%%%%%%%%%%%%%%%%%%%%%%%%%%%%%%%%%%%%%%%%%%%%%%%%%%
\subsection{\label{ssec:cs}Cross section}

The cross section was calculated from the experimental yields as 
\begin{equation}
\frac{d\sigma}{d\Omega} = \frac{A}{\left(\rho x\right) N_{\rm A}} \cdot \frac{N_K}{N_{\rm beam}} \cdot \frac{1}{\varepsilon_{\rm exp}\; d\Omega},
\label{eq:cs}
\end{equation}
where $A$ is the atomic mass of target hydrogen, $\rho x$ the target mass thickness, $N_{\rm A}$ Avogadro's number, $N_{\rm beam}$ the scaler counts of the beam trigger, $N_K$ the number of $(\pi, K)$ events,  $\varepsilon_{\rm exp}$ the total experimental efficiency, and $d\Omega$ the solid angle of SKS. Table~\ref{tab:eff} is a list of the experimental efficiency factors which consists of the beam normalization, detection efficiency, analysis efficiency, and other factors. Some factors depended on various experimental conditions, e.g., beam intensity, scattering angle, and momentum, which were taken into account.

\begin{table}[tb] %-----------------------------------------------------------
\caption{\label{tab:eff} List of experimental efficiency factors and their typical values in 2012 data.}
\begin{ruledtabular}
\begin{tabular}{ll}
 Efficiency                       & Typical value (\%) \\ \hline
 Beam normalization factor  &  90.2 $\pm$ 1.9 \\
 BC1,2 efficiency              &  85.0 $\pm$ 0.5 \\
 BC3,4 efficiency              &  99.1 $\pm$ 0.3 \\
 Beam spectrometer tracking efficiency  &  98.2 $\pm$ 0.3 \\
 Single track ratio               &  94.3 $\pm$ 0.3 \\
 SDC1,2 efficiency             &  97.4 $\pm$ 0.2 \\
 SDC3,4 efficiency             &  94.6 $\pm$ 1.1 \\
 SKS tracking efficiency      &  97.0 $\pm$ 0.8 \\
 TOF efficiency                 &  99.6 $\pm$ 2.5 \\
 LC efficiency                   &  97.5 $\pm$ 2.4 \\
 AC overkill factor              &  91.8 $\pm$ 2.1 \\
 PID efficiency for kaon       &  95.5 $\pm$ 2.0 \\
 Vertex cut efficiency         &  84.8 $\pm$ 1.0 \\
 Kaon decay factor             &  48.3 $\pm$ 0.4 \footnote{A representative value in case of a 1-GeV/$c$ momentum and a 5-m path length.} \\
 $K^-$ absorption factor      &  91.1 $\pm$ 1.2 \\
 Data acquisition efficiency   &  76.9 $\pm$ 0.5 \\
 Matrix trigger efficiency      &  98.6 $\pm$ 1.4 \\ \hline
 Total efficiency                &  15.1 $\pm$ 0.9 \\ 
\end{tabular}
\end{ruledtabular}
\end{table} %-----------------------------------------------------------

The beam normalization factor represents a fraction of the effective pion number in the beam trigger. Electrons in the beam were rejected by GC at trigger level. However, muons, which are decay products of pions, can not be separated from pions. The muon contamination rate was estimated to be $3.0 \pm 2.0\%$ by a Monte Carlo (MC) simulation using {\sc decay-turtle} \cite{Turtle}. The error represents the systematic uncertainty in the simulation. In previous experiments at KEK-PS \cite{MuonContami}, the muon contamination rate was measured and agreed with a {\sc decay-turtle} simulation within 2\%. The accidental coincidence rate between BH1 and BH2 was estimated to be $3.2 \pm 0.3\%$ by using time-of-flight spectra with the beam trigger. Considering the target vessel size, we applied the beam profile cut whose efficiency was typically $96.1 \pm 0.3\%$. In total, the beam normalization factor was $90.2 \pm 1.9\%$. 

The local straight-line tracking efficiencies of BC1,2, BC3,4, SDC1,2, and SDC3,4 were typically $85.0 \pm 0.5\%$, $99.1 \pm 0.3\%$, $97.4 \pm 0.2\%$, and $94.6 \pm 1.1\%$, respectively. The beam spectrometer tracking efficiency was $98.2 \pm 0.3\%$ and the single beam track fraction was $94.3 \pm 0.3\%$. The SKS tracking efficiency was estimated by using scattered proton events contaminating the data set of the $\pi^{\pm}p \to K^+\Sigma^{\pm}$ reaction because protons are free from an effect of decay in flight. Since the efficiency slightly depends on the incident angle to SKS, it was estimated angle by angle. A typical value was $97.0 \pm 0.8\%$. 

The TOF and LC efficiencies were estimated to be $99.6 \pm 2.5\%$ and $97.5 \pm 2.4\%$, respectively, 
by means of controlled data obtained by the trigger without each detector. The AC overkill rate was estimated to be $8.2 \pm 2.1\%$, which was caused by two factors: one was accidental coincidence ($3.3 \pm 0.6\%$) which was calculated by the AC single rate of 200 kHz and the trigger coincidence width; the other was induced by $\delta$ rays ($4.9 \pm 2.0\%$), which was estimated by using scattered proton events with the trigger without $\overline{\rm AC}$.

As described in Sec.~\ref{ssec:mm}, scattered kaons were identified by calculating $M_{\rm scat}^2$ (Fig.~\ref{fig:pid}). The kaon identification efficiency slightly depended on the momentum, which was taken into account. The typical efficiency was $95.5 \pm 2.0\%$. The uncertainty was due to the ambiguity of the low-mass tail of the kaon peak. 

The vertex cut efficiency was obtained by subtracting the empty-target data from the LH$_2$-target data (Fig.~\ref{fig:vtx}). The vertex cut efficiency strongly depends on the scattering angle because of the poor vertex resolution at small angle. Therefore, it was calculated angle by angle. The averaged value was $84.8 \pm 1.0\%$. The remaining background events from surrounding materials were estimated to be less than 3\% in the selected region. 

The kaon decay rate was corrected event by event using the momentum and the flight path length. Some kaons which decayed after passing through SDC4 fired AC or escaped from the acceptance. The probability of these leakages from the $(\pi, K)$ trigger was evaluated by a simulation. A typical kaon decay correction factor was $48.3 \pm 0.4\%$ in case of a 1-GeV/$c$ momentum and a 5-m path length. 

The $K^-N$ inelastic cross section is approximately 20 mb around 1 GeV/$c$ \cite{PDG2012}. The $K^-$ absorption rate was estimated to be $8.9 \pm 1.2\%$ by using a MC simulation.

A typical data-acquisition efficiency was measured to be $76.9 \pm 0.5\%$. The matrix-coincidence trigger efficiency of $98.6 \pm 1.4\%$ was obtained from the controlled data acquired by the trigger without the matrix coincidence.

Summarizing the efficiency factors described above, the overall efficiency factor was calculated event by event. A typical value was 15.1\% and a typical uncertainty was 0.9\%. Hence, the relative systematic uncertainty caused by the efficiency correction was estimated to be 6\%. 

The solid angle of SKS was calculated with a MC simulation as a function of the scattering angle and the momentum. The beam profile and the reaction vertex point were taken into account in the simulation. The typical uncertainty of 1\% was due to the statistical one in the simulation. 

\begin{figure}[tb] %-----------------------------------------------------------
\includegraphics[width=\linewidth]{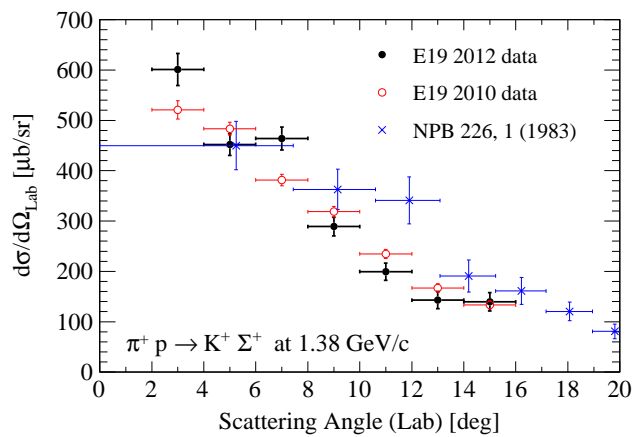}
\caption{\label{fig:sigma2}(Color online) Angular distribution for the $\pi^+p\to K^+\Sigma^+$ reaction at 1.38 GeV/$c$. The black solid circles and the red open circles are the present E19 data in 2012 and 2010, respectively. The blue crosses are the data from Candlin {\it et al.} \cite{Candlin1983}, converted from the center-of-mass frame to the laboratory one. The scattering angle is defined as the one between the outgoing kaon and the incoming pion. The quoted errors are statistical only.}
\end{figure}

Finally the total systematic uncertainty on the cross section was estimated to be 7\%, summing 6\% from the efficiency and 1\% from the acceptance. The validity of the efficiency and the acceptance correction was examined by using the known $\Sigma$ production cross sections. Figure~\ref{fig:sigma2} shows the differential cross sections of the $\pi^+p \to K^+\Sigma^+$ reaction at 1.38 GeV/$c$. Both of the present data sets taken in 2010 and 2012 are in good agreement with the previous experimental data \cite{Candlin1983}.

 %-----------------------------------------------------------

%%%%%%%%%%%%%%%%%%%%%%%%%%%%%%%%%%%%%%%%%%%%%%%%%%%%%%%%%%%%%%%%%%%%%%%%%%%%%%%%%%%%%%
%%%                                         RESULTS
%%%%%%%%%%%%%%%%%%%%%%%%%%%%%%%%%%%%%%%%%%%%%%%%%%%%%%%%%%%%%%%%%%%%%%%%%%%%%%%%%%%%%%
\section{\label{sec:result}Results}

\begin{figure}[tb] %-----------------------------------------------------------
\includegraphics[width=\linewidth]{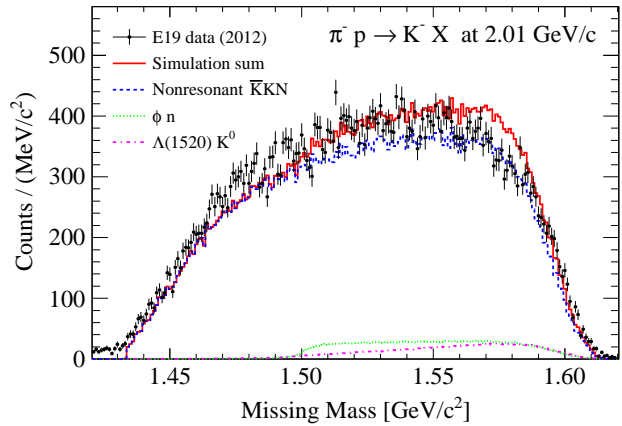}
\caption{\label{fig:mm1}(Color online) Missing mass spectrum of the $\pi^-p\to K^-X$ reaction at 2.01 GeV/$c$. The experimental data are indicated by black points with statistical errors. The red histogram represents the total background shape obtained by a MC simulation including three processes: nonresonant $\bar{K}KN$ (blue dashed), $\phi$-intermediated $K^-K^+n$ (green dotted) and $\Lambda(1520)$-intermediated $K^-K^0p$ (magenta dashed-dotted). The scale of the nonresonant components is normalized to the experimental data.}
\end{figure}

Figure \ref{fig:mm1} shows the missing mass spectrum of the $\pi^-p \to K^-X$ reaction at 2.01 GeV/$c$ at scattering angles from 2$^{\circ}$ to 15$^{\circ}$. The data are indicated as points with error bars. The spectrum is structureless and no clear peak was observed. 

In the present reaction, several background processes are associated with the $\Theta^+$ production. The $\pi^-p \to \bar{K}KN$ reaction at 1.8--2.2 GeV/$c$ was reported in Ref.~\cite{Dahl1967} and we considered following three processes as the main background: 
\begin{eqnarray}
\pi^- p & \to & \phi \; n \to K^- K^+ n , \label{eq:phi} \\
\pi^- p & \to & \Lambda(1520) \; K^0 \to K^- K^0 p , \label{eq:lambda} \\
\pi^- p & \to & K^- K^+ n \ {\rm or} \ K^- K^0 p \ \mbox{(nonresonant)}. \label{eq:nonres} 
\end{eqnarray}
Since other higher excited $\Lambda^*$ and $\Sigma^*$ resonances were not observed in \cite{Dahl1967}, we assumed the cross section is small and neglected the contribution. 

The background shape was reproduced by a MC simulation taking account of the reactions (\ref{eq:phi}--\ref{eq:nonres}). A background event was originated from a $K^-$ in the three-body final state detected in the spectrometer acceptance. The cross section and angular distribution of the $\phi$ and $\Lambda(1520)$ productions were taken from Refs.~\cite{Dahl1967,Courant1977}. Since there was no reliable information on the nonresonant cross section, the scale of the nonresonant contribution was normalized to the present experimental data. The simulated background spectra are overlaid as histograms in Fig.~\ref{fig:mm1}. Note that these background processes do not make any sharp structure in the missing mass spectrum.

 %-----------------------------------------------------------

\begin{figure}[tb] %-----------------------------------------------------------
\includegraphics[width=\linewidth]{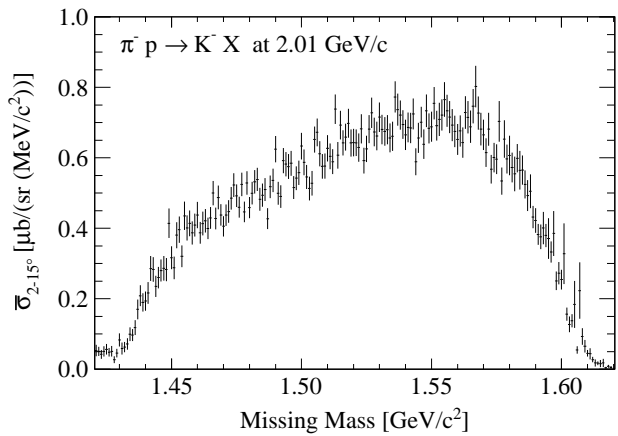}
\caption{\label{fig:mm2} Missing mass spectrum of the $\pi^-p\to K^-X$ reaction at 2.01 GeV/$c$. The ordinate represents the differential cross section averaged over 2--15$^{\circ}$ in the laboratory frame. The quoted errors are statistical.}
\end{figure} %-----------------------------------------------------------

\begin{figure}[tb] %-----------------------------------------------------------
\includegraphics[width=\linewidth]{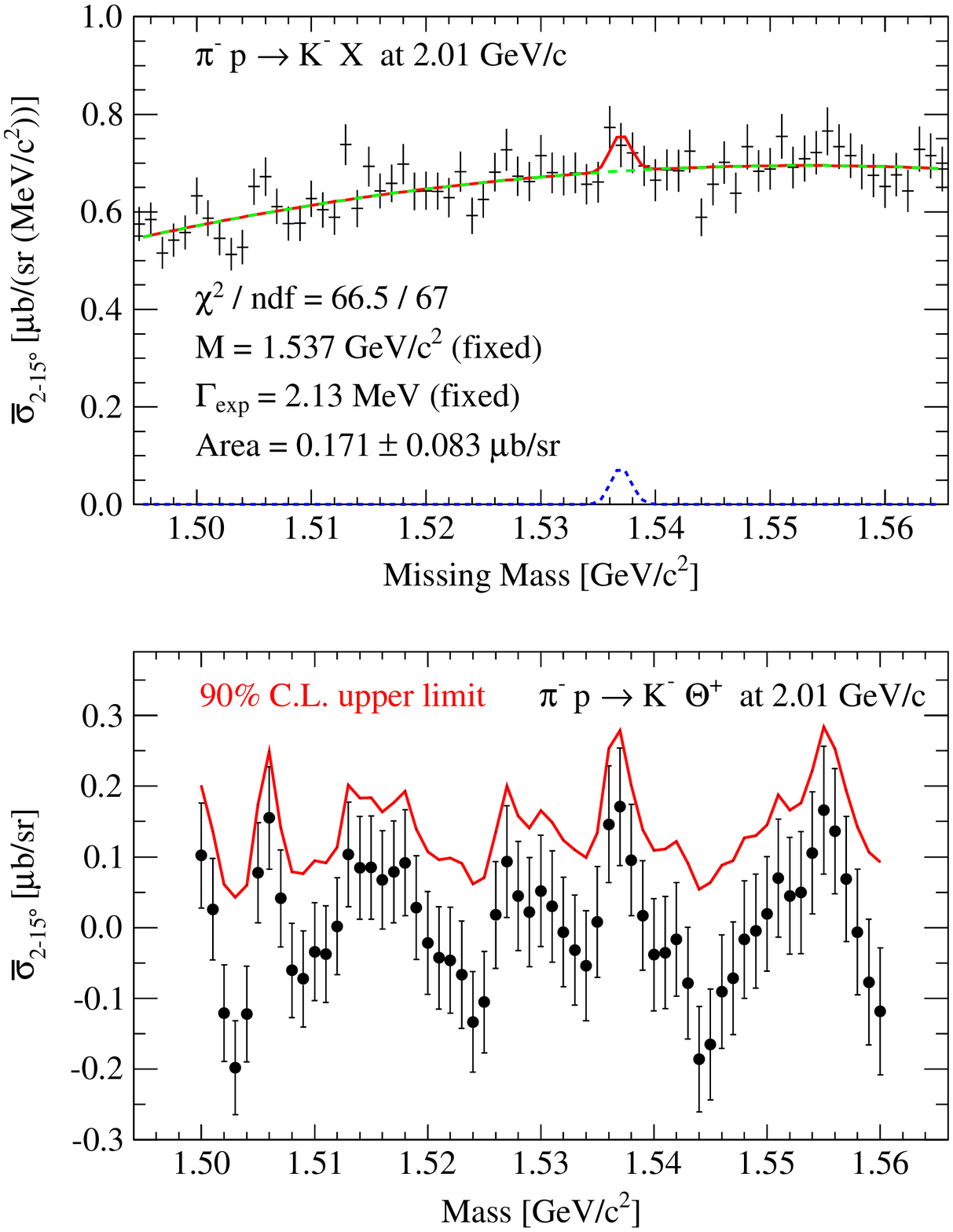}
\caption{\label{fig:ultheta}(Color online) (Top) Missing mass spectrum of the $\pi^-p\to K^-X$ reaction at 2.01 GeV/$c$ in the mass region of the $\Theta^+$ search. The quoted errors of the data are statistical. A fitting result at a mass of 1537 MeV/$c^2$ is also shown. The fitting function (red solid) is a second-order polynomial (green dashed) with a Gaussian peak (blue dotted) whose width is fixed by the experimental resolution of 2.13 MeV. (Bottom) Allowed signal yields for each mass. The error bars denote the statistical uncertainty. The red line indicates the upper limit at the 90\% confidence level.}
\end{figure} %-----------------------------------------------------------

Figure \ref{fig:mm2} shows the missing mass spectrum of the $\pi^-p \to K^-X$ reaction at 2.01 GeV/$c$ where the experimental efficiency and the acceptance were corrected; the ordinate represents the differential cross section averaged over 2$^{\circ}$ to 15$^{\circ}$ in the laboratory frame. We evaluated the upper limits on the $\Theta^+$ production cross section. As shown in the top figure of Fig.~\ref{fig:ultheta}, we fitted the spectrum with a background using a second-order polynomial function and a Gaussian peak with a width of 2.13 MeV (FWHM) which was the expected experimental resolution. The natural width for the $\Theta^+$ was ignored in the fitting. In addition, the error bar at each point indicates only the statistical uncertainty because the systematic uncertainty which came from the efficiency and acceptance correction was almost common within the local mass range of a few MeV/$c^2$. The systematic uncertainty is discussed later. Therefore the fitting provides an estimation of the effect from statistical fluctuation. The fitting was repeated for every assumed peak position from 1500 to 1560 MeV/$c^2$ with 1-MeV steps. Then the cross section was calculated from the area of the Gaussian function. The bottom figure of Fig.~\ref{fig:ultheta} shows the results and the upper limits at the 90\% confidence level. In the confidence level estimation, we assumed the Gaussian approximation where an unphysical region corresponding to negative cross section was excluded. The 90\% C.L. upper limit on the differential cross section averaged over 2--15$^{\circ}$ was derived to be at most 0.28 $\mu$b/sr in the mass region of 1500--1560 MeV/$c^2$. 

We investigated the influence upon the result from some systematic uncertainties. First, the result was affected by the cross section uncertainty of 7\% described in Sec.~\ref{ssec:cs}. Second, the mass resolution uncertainty of $\pm$0.15 MeV, which was described in Sec.~\ref{ssec:mr}, caused 4.2\% influence upon the upper limit. Finally, we examined an uncertainty due to the background shape. We applied a third-order polynomial function instead of the second-order one and found the influence negligibly small. As a whole, the systematic uncertainties were controlled within 10\% and did not have much influence upon the upper limits. 

As for the 2010 data \cite{Shirotori2012} taken at 1.92-GeV/$c$ beam momentum, we reevaluated the upper limits using the updated experimental resolution of 1.72 MeV described in Sec.~\ref{ssec:mr}. The upper limits were derived to be at most 0.28 $\mu$b/sr in the mass region of 1510--1550 MeV/$c^2$. 

Combining the 2010 and 2012 data, we have found the upper limits on the $\Theta^+$ production cross section to be less than 0.28 $\mu$b/sr both at 1.92 and 2.01 GeV/$c$. These are an order of magnitude lower than the previous E522 experimental result of 2.9 $\mu$b/sr \cite{Miwa2006}. We conclude that the bump structure observed in the E522 experiment was not a sign of $\Theta^+$ (and the authors did not claim so). Furthermore, the obtained upper limits are extraordinarily small as a hadronic production cross section. We quantitatively discuss it in the next section.

%%%%%%%%%%%%%%%%%%%%%%%%%%%%%%%%%%%%%%%%%%%%%%%%%%%%%%%%%%%%%%%%%%%%%%%%%%%%%%%%%%%%%%
%%%                                       DISCUSSION
%%%%%%%%%%%%%%%%%%%%%%%%%%%%%%%%%%%%%%%%%%%%%%%%%%%%%%%%%%%%%%%%%%%%%%%%%%%%%%%%%%%%%%
\section{\label{sec:disc}Discussion}

In this section, we discuss the constraint on the existence of the $\Theta^+$ focusing on its decay width. Theoretical calculations for the meson-induced $\Theta^+$ productions have been studied in Refs.~\cite{Oh2004-1,Hyodo2012,Liu2003,Hyodo2004,Oh2004-2,Ko2004,PKo2005,Hyodo2005}, where the authors adopted an effective interaction Lagrangian approach with several reaction mechanisms and different frameworks. In the $\pi^-p \to K^-\Theta^+$ reaction for the isosinglet $\Theta^+$, $s$- or $t$-channel diagram or two-meson coupling is allowed at tree level. Non-observation of $\Theta^+$ in the $K$-induced reaction \cite{Miwa2008} implied that the $t$-channel process, where the $K^{\ast}$ vector meson is exchanged, is quite small according to a preceding theoretical calculation \cite{Oh2004-2}. Two meson couplings of $N\pi K\Theta$ were studied in \cite{Hyodo2005}. Non-observation of $\Theta^+$ in both the $\pi$- and $K$-induced reactions \cite{Miwa2006,Miwa2008} resulted in the smallness of the two-meson coupling. Thus, the $s$-channel contribution seems to be dominant in the $\pi^-p \to K^-\Theta^+$ reaction. 

Recently, Hyodo {\it et al.} have published a comprehensive calculation \cite{Hyodo2012} which can be directly compared to our experimental result. They considered only the nucleon pole term which corresponds to the $s$-channel diagram in the $\pi^-p \to K^-\Theta^+$ reaction. Their calculation was performed for the isosinglet $\Theta^+$ with $J^P = 1/2^{\pm}$ and $3/2^{\pm}$ cases. They introduced two schemes for the Yukawa couplings, namely pseudoscalar (PS) and pseudovector (PV) schemes. They also introduced two types of form factors, namely static ($F_s$) and covariant ($F_c$) types, to reflect the finite size of the hadrons. Theoretical parameters were determined based on the known hadron reactions except for the unknown $KN\Theta$ coupling constant. Note that the $KN\Theta$ coupling constant corresponds to the $\Theta^+$ width. Since the amplitude for the $s$-channel diagram is proportional to the $KN\Theta$ coupling constant, the cross section ($\sigma_{\Theta}$) is simply proportional to the width of $\Theta^+$ ($\Gamma_{\Theta}$); 
\begin{equation}
\frac{d\sigma_{\Theta}}{d\Omega} = k_{C, F} \left( p_{\pi}, m_{\Theta} \right) \; \Gamma_{\Theta} ,
\label{eq:width}
\end{equation}
where the coefficient $k$ is obtained in each coupling scheme ($C$), PS or PV, and form factor ($F$), $F_s$ or $F_c$. $k$ is also a function of the incident momentum, $p_{\pi}$, and the $\Theta^+$ mass, $m_{\Theta}$. The differential cross section was calculated at the incident momenta of 1.92 and 2.00 GeV/$c$ and at the $\Theta^+$ mass every 10 MeV in the range 1510--1550 MeV/$c^2$. 

From the present experimental results, we obtained two structureless missing-mass spectra at 1.92 and 2.01 GeV/$c$. We simultaneously fitted these spectra with respect to a common width parameter, $\Gamma_{\Theta}$, which is related to the cross section according to Eq.~(\ref{eq:width}). The Breit-Wigner distribution smeared by the experimental resolution was used as a signal function. The experimental resolution was fixed at 1.72 and 2.13 MeV (FWHM) for the 1.92- and 2.01-GeV/$c$ data, respectively. The signal cross section was constrained by Eq.~(\ref{eq:width}). We allowed both positive and negative cross sections in fitting the spectra. In case of the negative cross section, the signal function was a Gaussian with the experimental resolution and negative height. As was done in Sec.~\ref{sec:result}, second-order polynomial functions were used as the background shape. The fitting result was obtained at each mass. The 90\% C.L. upper limits were estimated assuming the parabolic error and the Gaussian approximation in the same manner as described in Sec.~\ref{sec:result}.

\begin{figure}[tb] %-----------------------------------------------------------
\includegraphics[width=\linewidth]{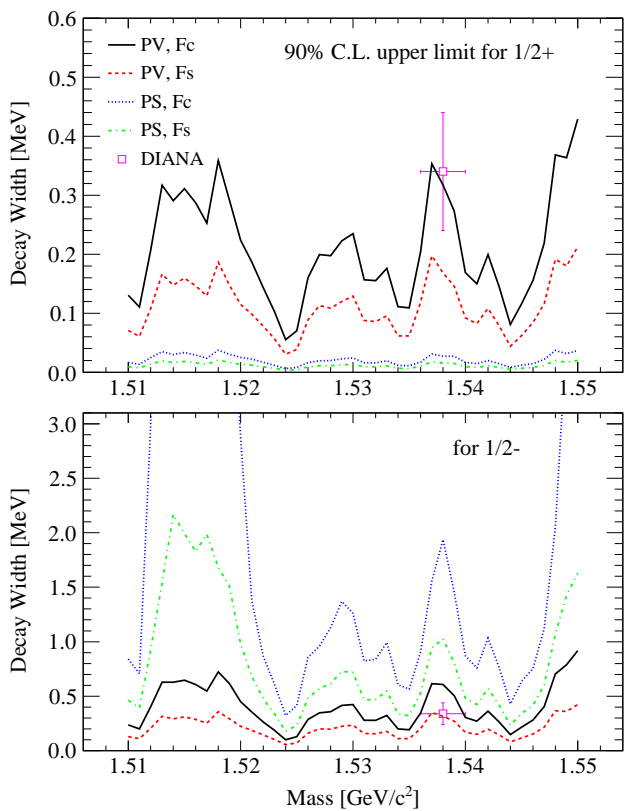}
\caption{\label{fig:width}(Color online) 90\% C.L. upper limit on the $\Theta^+$ decay width for the spin-parity of $1/2^+$ (top) and $1/2^-$ (bottom) case. Each line indicates the different theoretical treatments: pseudoscalar (PS) and pseudovector (PV) couplings, and static ($F_s$) and covariant ($F_c$) types of form factors. The DIANA result \cite{Diana2014} is also indicated by an open square (magenta).}
\end{figure} %-----------------------------------------------------------

Figure~\ref{fig:width} shows the obtained upper limits on the $\Theta^+$ decay width for each theoretical scheme and spin-parity. Here we considered the $1/2^{\pm}$ cases. The $3/2^{\pm}$ cases are highly disfavored in \cite{Hyodo2012} since the width derived from the previous experiments \cite{Miwa2006,Miwa2008} becomes too narrow. Because the difference among each scheme is a theoretical uncertainty, we took the most conservative one, where the obtained result gives the largest upper limit. In the $1/2^+$ case (top figure of Fig.~\ref{fig:width}), the PV scheme with $F_c$ form factor gives the largest upper limits. The upper limits of the decay width are less than 0.36 MeV in almost the entire mass region of 1510--1550 MeV/$c^2$. On the other hand, the $1/2^-$ case (bottom figure of Fig.~\ref{fig:width}) shows relatively larger width than the $1/2^+$ case. This can be understood by the partial wave of the $\Theta \to KN$ decay. $\Theta^+$ decays in $s(p)$-wave in the $1/2^-(1/2^+)$ case. In general, the decay width is smaller for higher wave with respect to the same coupling constant. In the $1/2^-$ case, the PS scheme with $F_c$ form factor gives the largest upper limits. The upper limits of the decay width are less than 1.9 MeV in the mass region around 1530 or 1540 MeV/$c^2$, whereas the sensitivity is not enough outside the range. 

We investigated the influence upon this result from the systematic uncertainty in the fitting. The missing mass has a systematic uncertainty due to the mass scale calibration. As was described in Sec.~\ref{ssec:msc}, the mass scale uncertainties were estimated to be 1.7 and 1.4 MeV in the 2010 and 2012 data, respectively. Due to this uncertainty, the upper limits could vary by $\pm$30\% and $^{+10}_{-30}$\% in the $1/2^+$ and $1/2^-$ cases, respectively. This was the dominant uncertainty in this fitting. 

Finally, the obtained upper limits on the $\Theta^+$ width are compared to other experimental results. From the viewpoint of hadron structure, for $\Theta^+$ with $1/2^-$, which decays in $s$-wave, it is difficult to explain the extraordinarily narrow width. We discuss $\Theta^+$ with $1/2^+$ next. We derived the upper limits on the width of less than 0.36 MeV in the possible $\Theta^+$ mass region in the most conservative case. Our limits are more stringent than both the old $K^+d$ scattering data \cite{Arndt2003,Haidenbauer2003,Cahn2004,Sibirtsev2004,Gibbs2004}, where the width was derived to be less than a few MeV, and Belle's upper limits, e.g., 0.64 MeV at 1539 MeV/$c^2$ \cite{Mizuk2006}. In Fig.~\ref{fig:width}, the DIANA result is also indicated. They claimed that the $\Theta^+$ was observed at 1538 $\pm$ 2 MeV/$c^2$ with the width of 0.34 $\pm$ 0.10 MeV \cite{Diana2014}. Our upper limits are comparable to their value. The consistency is subtle but our result does not completely contradict the DIANA claim.

%%%%%%%%%%%%%%%%%%%%%%%%%%%%%%%%%%%%%%%%%%%%%%%%%%%%%%%%%%%%%%%%%%%%%%%%%%%%%%%%%%%%%%
%%%                                       SUMMARY
%%%%%%%%%%%%%%%%%%%%%%%%%%%%%%%%%%%%%%%%%%%%%%%%%%%%%%%%%%%%%%%%%%%%%%%%%%%%%%%%%%%%%%
\section{\label{sec:sum}Summary}

We have searched for the pentaquark $\Theta^+$ via the $\pi^-p \to K^-X$ reaction at the K1.8 beam line in the J-PARC hadron facility. We acquired the experimental data at beam momenta of 1.92 and 2.01 GeV/$c$ with mass resolutions of 1.72 and 2.13 MeV (FWHM), respectively. No peak structure was observed in the missing mass spectra at scattering angles of 2--15$^{\circ}$ in the laboratory frame. The 90\% C.L. upper limits on the forward production cross section were found to be less than 0.28 $\mu$b/sr in both the 1.92- and 2.01-GeV/$c$ data for the possible $\Theta^+$ mass region. Combining with the theoretical calculation using the effective Lagrangian, where the cross section is proportional to the decay width of $\Theta^+$, constraints on the $\Theta^+$ decay width were evaluated. The 90\% C.L. upper limits on the decay width were derived to be less than 0.36 and 1.9 MeV for the $\Theta^+$ spin-parities of $1/2^+$ and $1/2^-$, respectively.

%%%%%%%%%%%%%%%%%%%%%%%%%%%%%%%%%%%%%%%%%%%%%%%%%%%%%%%%%%%%%%%%%%%%%%%%%%%%%%%%%%%%%%
\begin{acknowledgments}

We express our thanks to staffs of the J-PARC accelerator and the hadron beam line group for their outstanding efforts. We also acknowledge Tetsuo Hyodo for helpful theoretical discussions. This work was supported in part by Grants-in-Aid for Scientific Research (Nos. 17070001, 17070003, 17070006, and 22105512) from the Ministry of Education, Culture, Sports, Science and Technology, Japan. We acknowledge support from the National Research Foundation of Korea (No. 2010-0004752), the WCU program, the Center for Korean J-PARC Users, and the Ministry of Education, Science and Technology (Korea). We also thank KEKCC and SINET4. 

\end{acknowledgments}

%\nocite{*}


\end{document}